\RequirePackage{fix-cm}

\documentclass[smallcondensed]{svjour3}     
\smartqed  
\usepackage{graphicx}
\usepackage{amsfonts}
\usepackage{amsmath}
\usepackage{amssymb}
\usepackage{natbib}
\RequirePackage[colorlinks,citecolor=blue,urlcolor=blue]{hyperref}
\usepackage[ruled,vlined]{algorithm2e}
\usepackage{setspace}

\usepackage{geometry}
\def\bC{{\bf C}}
\def\bY{{\bf Y}}
\def\bz{{\bf z}}

\def\bPi{{\boldsymbol \Theta}}

\def\makeheadbox{\relax}

\begin{document}

\title{Bayesian Testing for Exogenous Partition Structures \\ in Stochastic Block Models
}

\author{Sirio Legramanti         \and Tommaso Rigon \and Daniele Durante  
}

\institute{Sirio Legramanti  \at
              Dept. of Decision Sciences, Bocconi University\\
              \email{sirio.legramanti@unibocconi.it}          
           \and
          Tommaso Rigon  \at
              Dept. of Statistical Sciences, Duke University
               \and
          Daniele Durante  \at
              Dept. of Decision Sciences  and  Institute for Data Science and Analytics, Bocconi University\\
}

\date{}

\maketitle
\vspace{-10pt}

\begin{abstract}
Network data often exhibit block structures characterized by clusters of nodes with similar patterns of edge formation.  When such relational data are complemented by additional information on exogenous node partitions, these sources of knowledge are typically included in the  model to supervise the cluster assignment mechanism or to improve inference on edge probabilities. Although these solutions are routinely implemented, there is a lack of formal approaches to test if a given external node partition is in line with the endogenous clustering structure encoding stochastic equivalence patterns among the nodes in the network. To fill this gap, we develop a formal Bayesian testing procedure which relies on the calculation of the Bayes factor between a stochastic block model with known grouping structure defined by the exogenous node partition and an infinite relational model that allows the endogenous clustering configurations  to be unknown, random and fully revealed by the block--connectivity patterns in the network. A simple Markov chain Monte Carlo method for computing the Bayes factor and quantifying uncertainty in the endogenous groups is proposed. This routine is evaluated in simulations and in an application to study exogenous equivalence structures in brain networks of Alzheimer's patients.

\end{abstract}

\section{Introduction}
\label{intro}
There is an extensive interest in learning grouping structures among the nodes in a network \citep[see, e.g.][]{fortunato2016}. Classical solutions to this problem focus on detecting community patterns via  algorithmic approaches that cluster  the nodes into groups characterized by a high number of edges within each community and comparatively few edges between the nodes in different communities \citep{newman_2004, blondel_2008, Fortunato_2010}. Despite being routinely implemented, these procedures do not rely on generative probabilistic models and, therefore, face difficulties when the focus is not just on point estimation, but also on hypothesis testing and uncertainty quantification.  This issue has motivated several  efforts towards developing model--based representations for inference on grouping structures, with the stochastic block model (\textsc{sbm}) \citep{Holland_1983, Nowicki_2001} providing the most notable contribution within this class. Such a statistical model expresses the edge probabilities as a function of the node assignments to groups and of block probabilities among such groups, thus allowing inference on more general block--connectivity patterns beyond community structures. The success of \textsc{sbm}s in different fields  has motivated various extensions \citep[e.g.][]{kemp_2006, air_2008, karrer_2011,geng_2019} and detailed theoretical studies on their  asymptotic properties \citep[e.g.][]{zhao2012, gao2018, van2018, ghosh2019}; see \citet{schmidt_2013, abbe_2017,lee2019} and the references therein for a comprehensive overview.

When node--specific attributes are available, the above block models have been generalized in different directions to incorporate such external information in the edge formation mechanism. Common proposals address this goal via the inclusion of nodal attributes within the generative model for the cluster assignments \citep[e.g.][]{tallberg_2004,white2016,newman2016,stanley2019}, or by defining the edge probabilities as a function of  block--specific parameters, as in classical \textsc{sbm}s, and of  pairwise similarity measures among node attributes  \citep[e.g.][]{mariadassou2010,choi2012,sweet2015,roy2019}. Such formulations are powerful approaches to assist  the cluster assignment mechanism and, typically, improve the estimation of the edge probabilities. However, when categorical node attributes are available, less attention has been paid to the development of formal Bayesian testing procedures to assess whether these exogenous partitions identified by an external  grouping variable are in line with the endogenous partition revealed by the block--connectivity behaviors in the network. For example, in structural brain network applications  it is often of interest to understand if exogenous anatomical partitions of the brain regions can accurately characterize the endogenous block structures of brain networks \citep[e.g.][]{sporns2013,faskowitz2018}. This goal could be partially addressed by the previously--mentioned models via inference on  the posterior distribution for the parameters regulating the effect of the node--specific attributes, but these formulations are prone to identifiability and computational issues. 

Motivated by the above discussion, we propose a formal yet simple Bayesian testing procedure to compare a stochastic block model with known grouping structure,  fixed according to a given exogenous node partition, and an infinite relational model \citep{kemp_2006} where the node assignments are unknown, random and modeled through a Chinese Restaurant Process (\textsc{crp}) prior  \citep{aldous1985}, which  allows the total number of non--empty clusters $H$ to be inferred. Such a Bayesian nonparametric representation allows flexible learning of the endogenous clustering configurations as revealed by the common connectivity behaviors within the network and, hence, provides a suitable reference model against which to assess the ability of a pre--specified exogenous partition to characterize the block--connectivity structures within the network. In a sense, our goal is related to those of \citet{bianconi2009} and \citet{peel2017}. However, such contributions compute, under a frequentist perspective, the entropy of a stochastic block model whose groups coincide with the external node partition and compare it with  the distribution of the entropies derived under the same network with grouping structure given by a random permutation of the exogenous node labels. Besides taking a Bayesian approach, our procedure quantifies proximities to endogenous block structures rather than studying departures from a random partition. This allows, as a byproduct, inference on node groupings supported by the data. In fact, leveraging the recent inference methods for Bayesian clustering \citep{wade2018} brought into the network field by \citet{legramanti2020}, we complement the results of the proposed testing procedures with an analysis of the credible balls for the grouping structure under the infinite relational model. 

In Section \ref{sec:1} we describe the proposed testing procedure, based on the calculation of the Bayes factor \citep[e.g.][]{kass1995} among the two competing  models, and discuss methods for uncertainty quantification on the inferred endogenous clustering. In Section \ref{sec:2}, we derive a collapsed Gibbs sampler to obtain samples from the posterior of the endogenous partition, thus allowing  Monte Carlo estimation of the marginal likelihood \citep{newton1994,raftery2006} required to compute the Bayes factor. As illustrated in simulations in Section~\ref{sec:3} and in an application to Alzheimer's brain networks in Section~\ref{sec:4}, the Gibbs sampler is also useful to perform inference on the endogenous groups. Codes to implement the proposed methods can be found at \url{https://github.com/danieledurante/TESTsbm}.

\vspace{-3pt}

\section{Model Formulation, Bayesian Testing and Inference}
\label{sec:1}
\vspace{-2pt}
\subsection{Endogenous and exogenous models}
\label{sec:1.2}
\vspace{-4pt}
Let $\bY$ denote an $n \times n$ symmetric adjacency matrix associated with an undirected binary network without self--loops, so that  $y_{vu}=y_{uv}=1$ if nodes $v=2, \ldots, n$ and $u=1, \ldots, v-1$ are connected, and $y_{vu}=y_{uv}=0$ otherwise. The absence of self--loops implies that the diagonal entries of $\bY$ are not considered for inference. Recalling our discussion in Section \ref{intro}, we consider a stochastic representation partitioning the nodes into exhaustive and non--overlapping groups, where nodes in the same group are characterized by equal patterns of edge formation. More specifically, let $\bz=(z_1, \ldots, z_n)^{\intercal} \in  \mathcal{Z}$ be the vector of cluster membership indicators for the $n$ nodes, with $\mathcal{Z}$ being the space of all the possible group assignments, so that $z_v=h$ if and only if the $v$th node belongs to the $h$th cluster. Letting $H$ be the number of non--empty groups in $\bz$, we denote with $\bPi$ the $H \times H$ symmetric matrix  of block probabilities with generic elements $\theta_{hk} \in (0,1)$ indexing the distribution of the edges between the nodes in cluster $h$  and those in cluster $k$. To characterize block--connectivity structures within the network, we assume  
\begin{eqnarray*}
(y_{vu} \mid z_v=h,z_u=k, \theta_{hk}) \sim \mbox{Bern}(\theta_{hk}),
\end{eqnarray*}
independently for each $v=2, \ldots, n$ and $u=1, \ldots, v-1$, with $\theta_{hk} \sim \mbox{Beta}(a,b)$, independently for every $h=1, \ldots, H$ and $k=1, \ldots, h$. This formulation recalls the classical Bayesian \textsc{sbm} specification \citep{Nowicki_2001} and leverages a stochastic equivalence property that relies on the conditional independence of the edges, whose distribution depends on the cluster membership of the associated nodes. Indeed, by marginalizing out the beta--distributed block probabilities, which are typically treated as nuisance parameters in the \textsc{sbm} \citep[e.g.][]{kemp_2006,schmidt_2013},  the likelihood for $\bY$ given $\bz$ is
\begin{equation}
p(\bY \mid \bz)=\prod_{h=1}^H\prod_{k=1}^h \frac{\mbox{B}(a+m_{hk}, b+ \bar{m}_{hk})}{\mbox{B}(a,b)},
\label{eq1}
\end{equation}
where $m_{hk}$ and $\bar{m}_{hk}$ denote the number of edges and non--edges among nodes in clusters $h$ and $k$, respectively, whereas $\mbox{B}(\cdot,\cdot)$ is the beta function. Expression \eqref{eq1} is derived by exploiting beta--binomial conjugacy, and, as we will clarify later in the article, is fundamental to compute Bayes factors and to develop a collapsed Gibbs sampler which updates only the endogenous cluster assignments while treating  the block probabilities as nuisance parameters.
Moreover, as is clear from  equation \eqref{eq1}, $p(\bY \mid {\bz}) $  is invariant under relabeling of the cluster indicators. Therefore $ p(\bY \mid \bz)$ is equal to  $p(\bY \mid \tilde{\bz}) $ for any relabeling $ \tilde{\bz} $ of $ \bz $, meaning that also the Bayes factors computed from these quantities are invariant under relabeling. Hence, in the rest of the paper, $ \bz $~will denote any element of the equivalence class of its relabelings, whereas $\mathcal{Z}$ will correspond to the space of all the partitions of $\{1,\dots,n\}$.

Recalling Section \ref{intro}, our goal is develop a formal Bayesian test to assess whether assuming $\bz$ as known and equal to an exogenous assignment vector ${\bz}^{*}$ produces an effective characterization of all the block structures in $\bY$, relative to what would be obtained by letting $\bz$ unknown, random and endogenously determined by the stochastic equivalence relations in $\bY$. The first hypothesized model  $ \mathcal{M}^* $ can be naturally represented via a \textsc{sbm} as in \eqref{eq1} with a fixed and known exogenous partition $\bz^*$, whereas the second model $ \mathcal{M}$ requires a flexible prior distribution for the  indicators $\bz=(z_1, \ldots, z_n)^{\intercal}$ which is able to reveal the endogenous grouping structure induced by the block--connectivity patterns in $\bY$, without imposing strong parametric constraints.  A natural option would be to consider a Dirichlet--multinomial prior as in classical \textsc{sbm}s \citep{Nowicki_2001}, but such a specification requires the choice of the total number of groups, which is typically unknown. This issue is usually circumvented  by relying on \textsc{bic} metrics that require estimation of multiple \textsc{sbm}s \citep[e.g.][]{saldana_2017}. To avoid these computational costs and increase flexibility, we rely on a Bayesian nonparametric solution that induces a full--support prior on the total number $H$ of non--empty groups in $\bz$. This enables learning of $H$, which is not guaranteed to coincide with the number $H^*$ of non--empty groups in $\bz^*$. A widely used prior in the context of \textsc{sbm}s is the \textsc{crp}  \citep{aldous1985}, which leads to the so--called infinite relational model  \citep[][]{kemp_2006,schmidt_2013}. Under this prior  each group attracts new nodes in proportion to its size, and the formation of new groups depends only on the size of the network and on a tuning parameter $\alpha>0$. More specifically, under model $ \mathcal{M}$, we assume the following prior over cluster indicators for the $v$th node, given the memberships $\bz_{-v}=(z_1,\ldots ,z_{v-1},z_{v+1}, \ldots, z_{n})^{\intercal}$  of the others 
\begin{eqnarray}
\mbox{pr}(z_v=h \mid \bz_{-v}) = \begin{cases} \frac{n_{h,-v}}{n-1+\alpha}& \quad \text{if} \ \ h=1, \ldots, H_{-v}, \\ \frac{\alpha}{n-1+\alpha} & \quad \text{if} \ \ h=H_{-v}+1. \\ \end{cases} 
 \label{CRP}
\end{eqnarray}
In \eqref{CRP}, $H_{-v}$ is the number of non--empty groups in $\bz_{-v}$, the integer $n_{h,-v}$ is the total number of nodes in cluster $h$, excluding the $v$th one, whereas $\alpha >0$ denotes a concentration parameter controlling the expected number of non--empty clusters. The urn representation in equation \eqref{CRP} is induced by the joint probability mass function $p(\bz)=\alpha^H [\prod_{h=1}^H(n_h-1)!][\prod_{v=1}^n(v-1+\alpha)]^{-1}$, which shows that the \textsc{crp} is exchangeable. See also \citet{gershman2012} for an overview of \textsc{crp} priors.

\vspace{-5pt}

\subsection{Bayesian testing}
\label{sec:2.2}
\vspace{-5pt}
To compare the ability of the endogenous ($ \mathcal{M} $) and exogenous ($ \mathcal{M}^* $) formulations in characterizing the block structures  in $ \bY $, we define a formal Bayesian test relying on the Bayes factor. More specifically, assuming that the two competing models are equally likely a priori, i.e. $p(\mathcal{M})=p(\mathcal{M}^*)$, we compare $ \mathcal{M}$ against $\mathcal{M}^*$ via
\begin{equation}
{\mathcal{B}}_{\mathcal{M},\mathcal{M}^*}=\frac{p(\bY \mid \mathcal{M})}{p(\bY \mid \mathcal{M}^*)}=\frac{\sum_{\bz \in \mathcal{Z}} p(\bY \mid \bz) p(\bz)}{p(\bY \mid \bz^*)},
\label{eqBAY}
\end{equation}
where  $\sum_{\bz \in \mathcal{Z}} p(\bY \mid \bz) p(\bz)$ and $p(\bY \mid \bz^*)$ are the marginal likelihoods of $ \bY $ under $\mathcal{M}$ and $\mathcal{M}^* $. Recalling, e.g., \citet{kass1995}, equation \eqref{eqBAY} defines a formal Bayesian procedure to assess evidence against $\mathcal{M}^*$ relative to $\mathcal{M}$, with high values suggesting that the exogenous assignments in $\bz^*$ are not as effective in characterizing the endogenous block structures in $\bY$ as the posterior for $\bz$ under $\mathcal{M}$. Under the assumption that $p(\mathcal{M})=p(\mathcal{M}^*)$, the Bayes factor in \eqref{eqBAY} coincides with  the posterior odds $p(\mathcal{M} \mid \bY)/p(\mathcal{M}^* \mid \bY)$. When $p(\mathcal{M}) \neq p(\mathcal{M}^*)$, it suffices to rescale ${\mathcal{B}}_{\mathcal{M},\mathcal{M}^*}$ by $p(\mathcal{M})/p(\mathcal{M}^*)$ to assess posterior evidence against  $\mathcal{M}^*$ relative to $\mathcal{M}$.

To evaluate equation \eqref{eqBAY}, note that the quantity $p(\bY \mid \bz^*)$ can be computed by evaluating \eqref{eq1} at $\bz=\bz^*$. In contrast, model $\mathcal{M}$ requires the calculation of $p(\bY \mid \bz)$ and $p(\bz)$ for every $\bz \in  \mathcal{Z}$. Although both  quantities can be evaluated in closed form as discussed in Section \ref{sec:1.2}, this approach is computationally impractical due to the huge cardinality of the set $\mathcal{Z}$, thus requiring alternative strategies relying on Monte Carlo estimation of $p(\bY {\mid} \mathcal{M})  =\sum_{\bz {\in} \mathcal{Z}} p(\bY {\mid} \bz) p(\bz)$ via importance sampling methods. Here, we consider the harmonic mean approach  \citep{newton1994,raftery2006}, thus obtaining
\begin{eqnarray}
\hat{p}(\bY \mid \mathcal{M})= \left[\frac{1}{R} \sum_{r=1}^R \frac{1}{p(\bY \mid \bz^{(r)})} \right]^{-1},
 \label{harm}
\end{eqnarray}
where $\bz^{(1)}, \ldots, \bz^{(R)}$ are samples from the posterior distribution of $\bz$ and $p(\bY \mid \bz^{(r)})$ is the likelihood in \eqref{eq1} evaluated at $\bz=\bz^{(r)}$, for every $r=1, \ldots, R$. The harmonic mean approach is a consistent strategy to evaluate marginal likelihoods and, due to its simplicity, is widely implemented. Although recent refinements have been proposed to address some shortcomings of the harmonic estimate \citep[e.g.][]{lenk2009,pajor2017}, here we consider the original formula which is computationally more tractable and has proved stable in our simulations and applications. 
 
Leveraging equations \eqref{eq1} and  \eqref{harm}, our estimate of the Bayes factor in \eqref{eqBAY} is
\begin{eqnarray}
\hat{\mathcal{B}}_{\mathcal{M},\mathcal{M}^*}=\frac{\hat{p}(\bY \mid \mathcal{M})}{p(\bY \mid \mathcal{M}^*)}=\frac{\left[\frac{1}{R} \sum_{r=1}^R \prod_{h=1}^{H^{(r)}}\prod_{k=1}^h\frac{\mbox{\small B}(a,b)}{\mbox{\small B}(a+m^{(r)}_{hk}, b+ \bar{m}^{(r)}_{hk})} \right]^{-1}}{\prod_{h=1}^{H^*}\prod_{k=1}^h \frac{\mbox{\small B}(a+m^*_{hk}, b+ \bar{m}^*_{hk})}{\mbox{\small B}(a,b)}},
 \label{eqBAY_est}
\end{eqnarray}
where $m^{(r)}_{hk}$ and $\bar{m}^{(r)}_{hk}$ refer to the counts of edges and non--edges among nodes in groups $h$ and $k$ induced by the $r$th \textsc{mcmc} sample of $\bz$, whereas  $m^{*}_{hk}$ and $\bar{m}^{*}_{hk}$ denote the number of edges and non--edges among the nodes in clusters $h$ and $k$ induced by the exogenous assignments $\bz^*$. Finally, $H^{(r)}$ and $H^{*}$ are the total numbers of unique labels in  $\bz^{(r)}$ and $\bz^{*}$. Section \ref{sec:2} describes the collapsed Gibbs algorithm to sample the assignment vectors $\bz^{(1)}, \ldots, \bz^{(R)}$ from the posterior  $p(\bz \mid \bY)$ under model $\mathcal{M}$. These samples are required to compute \eqref{eqBAY_est} and, as discussed in Section \ref{sec:2.3}, also allow inference on the posterior distribution of the endogenous partitions.
 
\vspace{-5pt}

\subsection{Inference and uncertainty quantification on the endogenous partition}
\label{sec:2.3}
\vspace{-5pt}
When the Bayes factor discussed in Section \ref{sec:2.2} provides evidence in favor of model $\mathcal{M}$, it is of interest to study the posterior distribution of $\bz$ leveraging the Gibbs samples $\bz^{(1)}, \ldots, \bz^{(R)}$. Common  strategies address this goal by first computing the posterior co--clustering matrix $\bC$ with elements $c_{vu}=c_{uv}$ measuring the relative frequency of the Gibbs samples in which nodes $v=2, \ldots, n$ and $u=1, \ldots, v-1$ are in the same cluster, and then apply a standard clustering procedure to such a similarity matrix. However, this approach provides only a point estimate of $\bz$ and, hence, fails to quantify  posterior uncertainty. \citet{legramanti2020} recently covered this gap by adapting the novel inference methods for Bayesian clustering in \citet{wade2018} to the network field. These strategies rely on the variation of information (\textsc{vi}) metric, which quantifies distances between two partitions by  comparing their individual and joint entropies.

Under this framework, a point estimate $\hat{\bz}$ for $\bz$ coincides with that partition having the lowest  posterior averaged \textsc{vi} distance from all the other clusterings, whereas a $1-\delta$ credible ball around $\hat{\bz}$ is obtained by collecting all those partitions with a \textsc{vi} distance from $\hat{\bz}$ below a given threshold, with this threshold chosen to guarantee the smallest--size ball containing at least $1-\delta$ posterior probability. Such inference is useful to complement the results of the test  in Section~\ref{sec:2.2}. Namely, to get further reassurance about the output of the proposed test, we may also study whether the exogenous clustering $\bz^*$ is plausible under the posterior distribution for the endogenous partition $\bz$ by checking if $\bz^*$ lies inside the credible ball around $\hat{\bz}$.  Refer to \citet{wade2018,legramanti2020} and to the codes at \url{https://github.com/danieledurante/TESTsbm} for more details on the aforementioned inference methods and their  implementation.
 
Finally, although the block probabilities are integrated out, a plug--in estimate for these quantities can be easily obtained. Indeed, leveraging beta--binomial conjugacy, $(\theta_{hk} \mid \bY,\bz) \sim \mbox{Beta}(a+m_{hk}, b+\bar{m}_{hk}) $. Hence, a plug--in estimate of the block probabilities $\theta_{hk}$ for $h=1,\ldots,\hat{H}$ and $k =1,\ldots,h$ is
\begin{eqnarray*}
\hat{\theta}_{hk}=\mathbb{E}[{\theta_{hk} \mid \bY,\hat{\bz}]= \frac{a+\hat{m}_{hk}}{a+\hat{m}_{hk}+b+\hat{\bar{m}}_{hk}}},
\label{eq5}
\end{eqnarray*}
where $\hat{m}_{hk}$ and $\hat{\bar{m}}_{hk}$ denote the number of edges and non--edges between nodes in groups $h$ and $k$, respectively, induced by the posterior point estimate $\hat{\bz}$ of $\bz$. 

\section{Posterior Computation via Collapsed Gibbs Sampling}
\label{sec:2}
The posterior samples of $\bz$ under model \eqref{eq1} with \textsc{crp} prior \eqref{CRP} can be obtained via a simple collapsed Gibbs sampler which updates the group assignment of each node $v$ conditioned on those of the others by sampling from the full--conditional distribution $p(z_v \mid \bY, \bz_{-v})$ \citep{schmidt_2013}. By collapsing out the beta priors for the block probabilities, this procedure reduces the computational time in avoiding the updating of $\theta_{hk}$ for  $h=1, \ldots, H $ and $k=1, \ldots, h$, while  improving mixing \citep{neal2000}. 

Algorithm \ref{Algorithm} provides the detailed steps of one cycle of the Gibbs sampler. Note that since equation \eqref{eq1} is the joint probability for a large set of binary edges, manipulating this quantity within Algorithm  \ref{Algorithm} and in computing the Bayes factor in \eqref{eqBAY_est} may lead to practical difficulties due to the need to deal with quantities very close to zero. In these settings, we suggest to work with the logarithm, when possible, and to exploit the log--sum--exp identity $\log[ \sum_{i} \exp(\nu_i)]=\mbox{d}+\log[ \sum_{i} \exp(\nu_i-\mbox{d})]$, where $\mbox{d}$ usually coincides with $\max_i \nu_i$.

\begin{algorithm}[t]
 \For{$v=1, \ldots, n$}{
Update each $z_{v}$ conditionally on $\bz_{-v}$ and $\bY$ as follows
\begin{enumerate}
\item Remove node $v$ from the node set.
\item If no other node belongs to the cluster of $v$, such a cluster is removed.
\item Reorder  the cluster indices so that $1, \ldots, H_{-v}$ are non--empty and sample $z_{v}$\\ from the categorical variable with full--conditional probabilities
\begin{eqnarray*}
\mbox{pr}(z_{v}=h \mid \bY,\bz_{-v} )\propto \begin{cases} \frac{n_{h,-v}}{n-1+\alpha}p(\bY \mid z_{v}=h, \bz_{-v}), & \mbox{if} \  \ h=1, \ldots, H_{-v}, \\ \frac{\alpha}{n-1+\alpha}p(\bY \mid z_{v}=h, \bz_{-v}), &\mbox{if} \  \ h=H_{-v}+1,
\end{cases} 
\end{eqnarray*}
where $p(\bY \mid z_{v}=h, \bz_{-v})$ is computed as in \eqref{eq1} conditioned on $z_{v}=h$ and $ \bz_{-v}$.
\end{enumerate} }
  \Return $\bz=(z_{1}, \ldots, z_{n})^{\intercal}$ \\
    \caption{{\bf One step of  the  Gibbs sampler for $\bz$ under $\mathcal{M}$} \label{Algorithm}}
\end{algorithm}
 
\vspace{-3pt}

\section{Simulation Studies}
\label{sec:3}
\vspace{-2pt}

We consider an illustrative simulation to assess the performance of the new inference procedures presented in Section \ref{sec:1}, and to evaluate the ability of model $\mathcal{M}$ to recover underlying endogenous partition structures. Consistent with this goal, we simulate a symmetric binary adjacency matrix $\bY$ from a stochastic block model with $n=60$ nodes partitioned into $H_0=3$ groups of equal size. In particular, we let $\bz_0=(z_{1,0}=1, \ldots, z_{20,0}=1, z_{21,0}=2, \ldots, z_{40,0}=2, z_{41,0}=3, \ldots, z_{60,0}=3)^{\intercal}$, and simulate each $y_{vu}=y_{uv}$ for $v=2, \ldots, n, \ u=1, \ldots, v-1$ from a Bernoulli with probability $0.8$ if nodes $v$ and $u$ are in the same group, and $0.2$ otherwise.

\begin{figure*}
\centering
  \includegraphics[width=12cm]{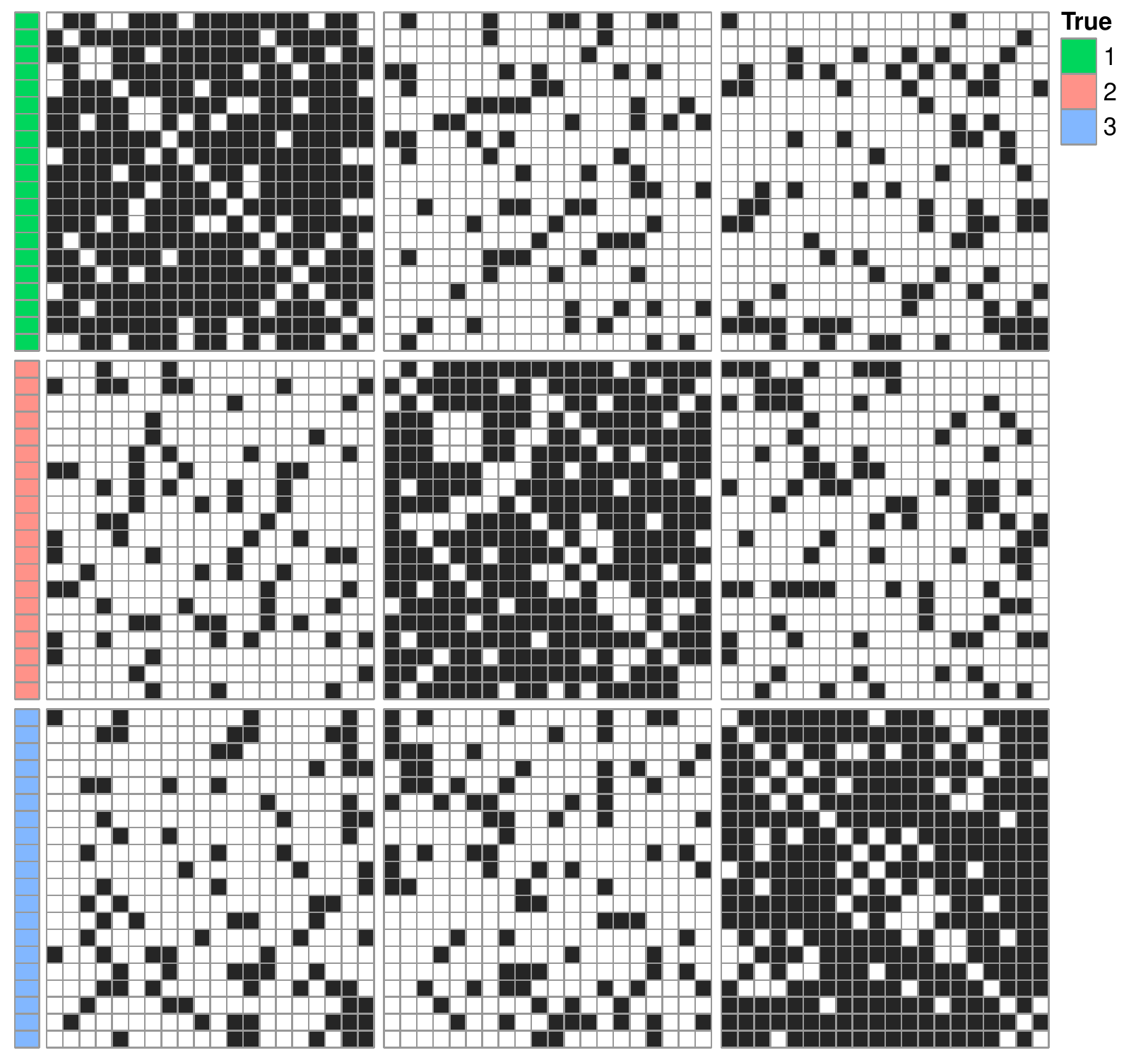}
\caption{Graphical representation of the simulated adjacency matrix $\bY$ partitioned in blocks according to the estimated endogenous assignments $\hat{\bz}$. Black and white cells denote edges and non--edges, respectively, whereas the first colored column represents the true  partition $\bz_0$.}
\label{fig:1}      
\end{figure*}

In performing posterior inference on the endogenous clustering structure under model $\mathcal{M}$, we set $a=b=1$ to induce a uniform prior on the block probabilities. This choice is theoretically supported \citep[e.g.][]{ghosh2019} and has been widely employed in routine implementations of \textsc{sbm}s \citep[][]{Nowicki_2001,kemp_2006,geng_2019}. As for $\alpha$ in prior \eqref{CRP}, we set it equal to~$1$ following default specifications of the \textsc{crp}, thus circumventing the need to include a hyper--prior which could affect mixing and convergence of Algorithm \ref{Algorithm}. Such a default specification has proved effective both in simulations and in applications, and we found the results robust to moderate changes in $\alpha$. For instance, setting $\alpha=0.1$ or $\alpha=10$ did not change the final conclusions of our testing procedures. 

\begin{figure*}
  \includegraphics[width=12.3cm]{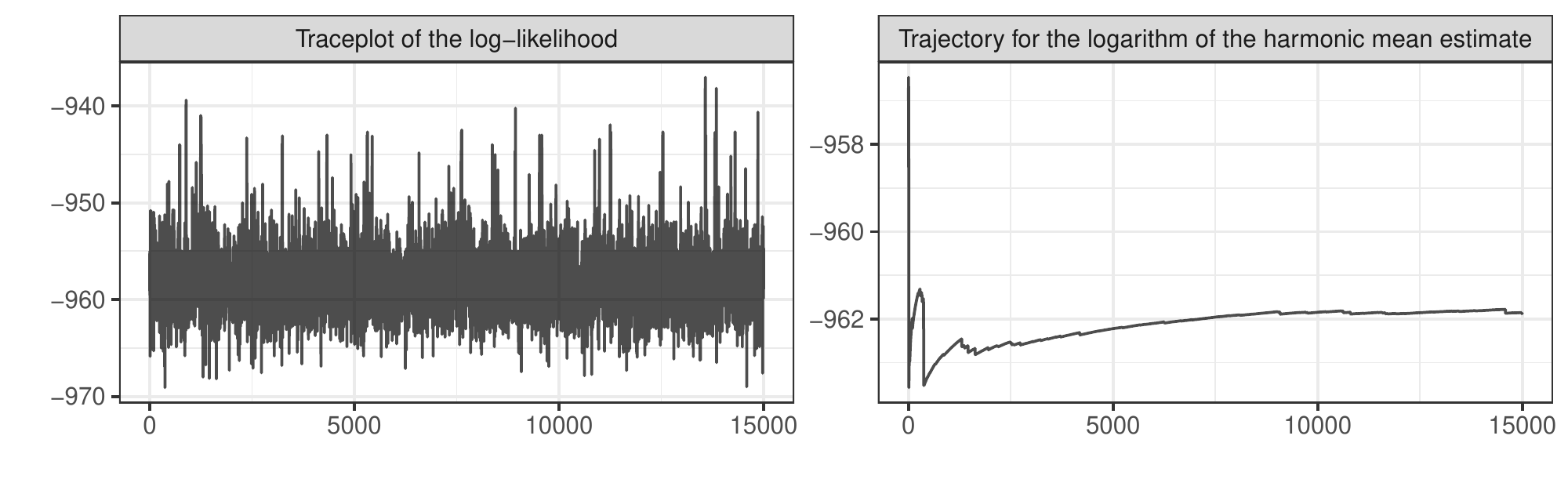}
\caption{\textsc{mcmc} diagnostics for the simulation study. Left: traceplot for the logarithm of the likelihood in equation  \eqref{eq1} computed at the \textsc{mcmc} samples of $\bz$ after burn--in. Right: trajectory of the logarithm of the harmonic mean estimate in equation \eqref{harm} for growing $R$.}
\label{fig:1.1}      
\end{figure*}

Figure \ref{fig:1} shows the simulated adjacency matrix $\bY$ paritioned in blocks according to the estimated  $\hat{\bz}$ under model $\mathcal{M}$. Such an estimate relies on $15000$ \textsc{mcmc} samples produced by Algorithm \ref{Algorithm}, after a burn--in of $2000$. As shown in Figure \ref{fig:1.1}, such settings are sufficient for good convergence and mixing according to the \textsc{mcmc} diagnostics of key measures for posterior inference, covering the traceplot of the log--likelihood in \eqref{eq1} under model $\mathcal{M}$ and the trajectory of the logarithm of the harmonic mean estimate for the associated marginal likelihood in \eqref{harm}. As is clear from the block partition of $\bY$ in Figure \ref{fig:1}, the posterior for $\bz$ under model $\mathcal{M}$ is able to concentrate around the true underlying endogenous partition and allows learning of the correct number of non--empty groups. These results support the use of $\mathcal{M}$ as a benchmark to test for differences between endogenous and exogenous partitions under the methods presented in Section  \ref{sec:1}. 

To assess the quality of such strategies, we consider four external assignment vectors $\bz_0, \bz_1, \bz_2$ and $\bz_3$ evaluated in Table \ref{tab:1}. In particular,  $\bz_0$ denotes the true generative partition, $\bz_1$ is obtained by a random permutation of the indices in $\bz_0$, while $\bz_2$ and $\bz_3$ define a refined and a coarsened partitioning of $\bz_0$, respectively, in which each cluster is either divided in two additional ones ($\bz_2$) or collapsed with others to form a single group ($\bz_3$). Due to this, we expect to obtain evidence in favor of the exogenous partition only in the scenario with $\bz^*=\bz_0$. Table \ref{tab:1} confirms our expectations when compared with the thresholds in \citet{kass1995}. Note that, although $\hat{\bz}=\bz_0$, we obtain a negative Bayes factor in the first scenario, which leads to a strong preference for model $\mathcal{M}^*$ relative to $\mathcal{M}$. Indeed, even if the point estimate $\hat{\bz}$ for $\bz$ under $\mathcal{M}$ exactly recovers $\bz_0$, there is still some amount of posterior uncertainty induced by the \textsc{crp} prior on $\bz$. On the contrary, $\mathcal{M}^*$ is defined in the first scenario by conditioning on the true underlying partition  with no uncertainty, thus providing a formulation much closer to the true data--generative mechanism relative to $\mathcal{M}$. All the remaining exogenous partitions $\bz_1, \bz_2$ and $\bz_3$ are, instead, not as effective as model $\mathcal{M}$ in characterizing the endogenous block structures within $\bY$. As expected, this is especially true for the random partition $(\bz_1)$, but also those obtained from refinements ($\bz_2$) or coarsening ($\bz_3$) operations on $\bz_0$ are not plausible according to the results of the tests. Such results confirm the ability of our procedures to provide accurate conclusions under various configuration of $\bz^*$. For instance, although the partition $\bz_2$ still leads to homogenous blocks in $\bY$, the additional refinements in $\bz_2$ provide an unnecessary addition of further groups which are  not required to characterize the block--connectivity patterns in $\bY$, thus leading the test to provide evidence in favor of $\mathcal{M}$ rather than $\mathcal{M}^*$ when $\bz^*=\bz_2$.

The \textsc{vi} distances between the estimated $\hat{\bz}$ under model $\mathcal{M}$ and the four exogenous partitions confirm the results of the tests. In particular, the only external assignment vector with a \textsc{vi} distance from $\hat{\bz}$ less than the estimated  $0.428$ threshold of the $95\%$ credible ball around $\hat{\bz}$ is $\bz_0$.

\begin{table*}
\centering
\caption{Results of our proposed procedure for testing to what extent four different exogenous partitions are as effective as the infinite relational model $\mathcal{M}$ in characterizing the endogenous block structures within $\bY$. The \textsc{vi} distance between the estimated partition $\hat{\bz}$ under the infinite relational model  and the exogenous ones is also displayed.}
\label{tab:1}
\begin{tabular}{lcccc}
\hline\noalign{\smallskip}
$\bz^*$ & $\bz_0$ (True) & $\bz_1$ (Random) & $\bz_2$   (Refined) & $\bz_3$ (Coarsened)  \\
\noalign{\smallskip}\hline\noalign{\smallskip}
$2 \log \hat{\mathcal{B}}_{\mathcal{M},\mathcal{M}^*}$ & $-5.17$ & $522.27$ & $25.68$ & $260.40$  \\
$\textsc{vi}(\hat{\bz},\bz^*)$ & $0.00$ & $3.16$ & $1.00$ & $0.67$\\
\noalign{\smallskip}\hline
\end{tabular}
\end{table*}

\vspace{-4pt}

\section{Application to Brain Networks of Alzheimer's Individuals}
\label{sec:4}
\vspace{-2pt}

\begin{figure*}
\centering
  \includegraphics[width=12.5cm]{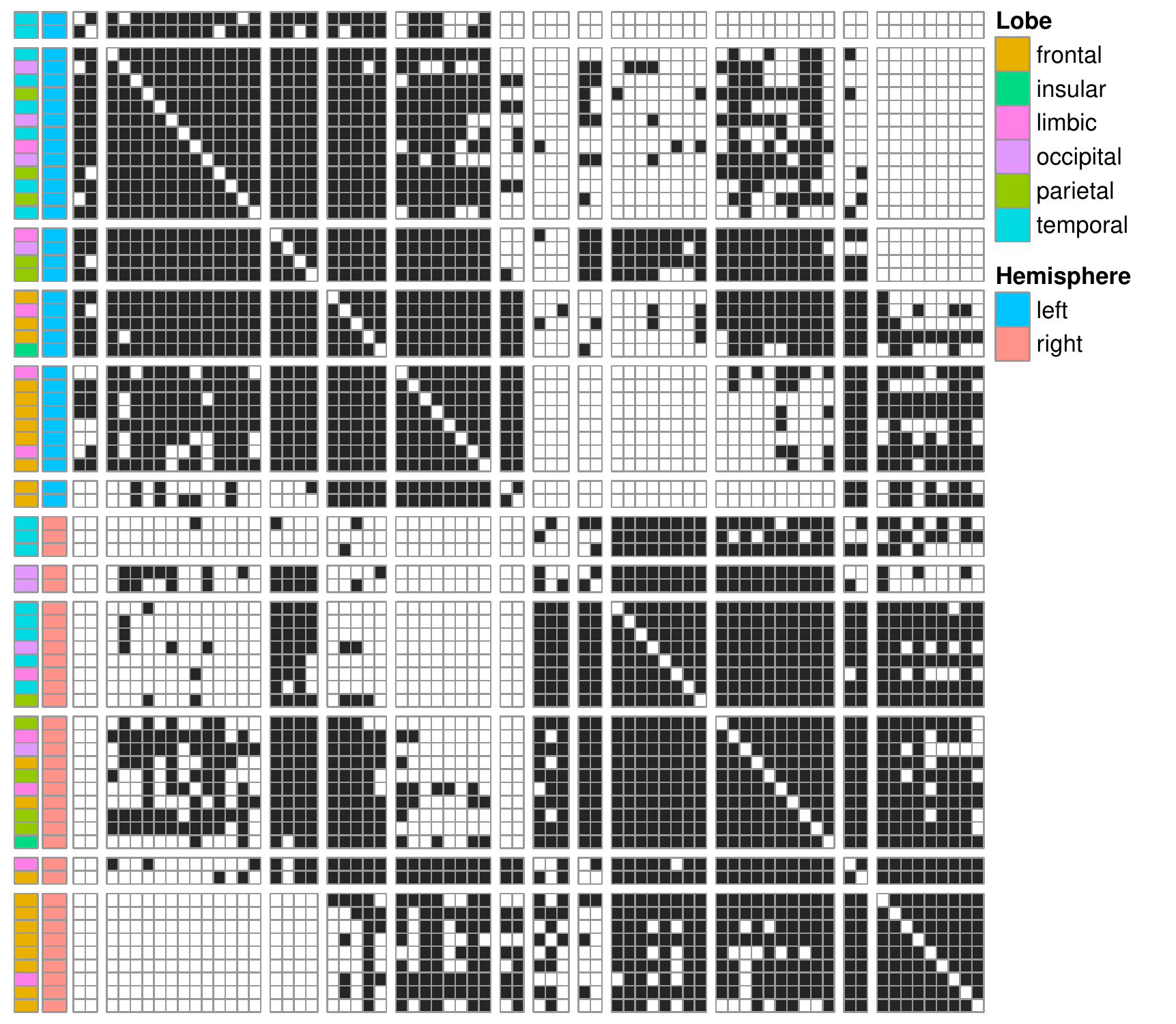}
\caption{Graphical illustration of a representative brain network $\bY$ for Alzheimer's individuals. Brain regions are re--ordered and partitioned in blocks according to the estimated endogenous assignments $\hat{\bz}$. Black and white cells denote edges and non--edges, respectively, whereas the first two colored columns represent the two exogenous anatomical brain partitions into hemispheres and lobes.}
\label{fig:2} 
\end{figure*}

There is an intensive research effort aimed at finding the sources of the Alzheimer's disease in human brain networks. Such an increasing interest is motivated by recent developments in  brain imaging technologies and by the constant growth of elderly population in the age interval mostly affected by Alzheimer's, thus making such a disease a major concern, both in terms of disability and mortality, especially for countries with longer life expectancy \citep{ashford2011handbook,ashford2011,stam2014}. Here, we focus on studying  structural brain networks encoding presence or absence of white matter fibers among anatomical regions in human brains. Such connectivity data have been a source of major interest in several recent studies mostly focused on  topological summary measures of Alzheimer's brains and on how these measures change as the disease progresses \citep[][]{daianu2013,sulaimany_2017,john2017,maartensson2018}. Instead, we consider a different perspective by studying the endogenous block structures in a representative Alzheimer's brain network, while assessing whether exogenous region partitions of interest can effectively characterize the block structures of such a network.

\begin{figure*}
  \includegraphics[width=12.3cm]{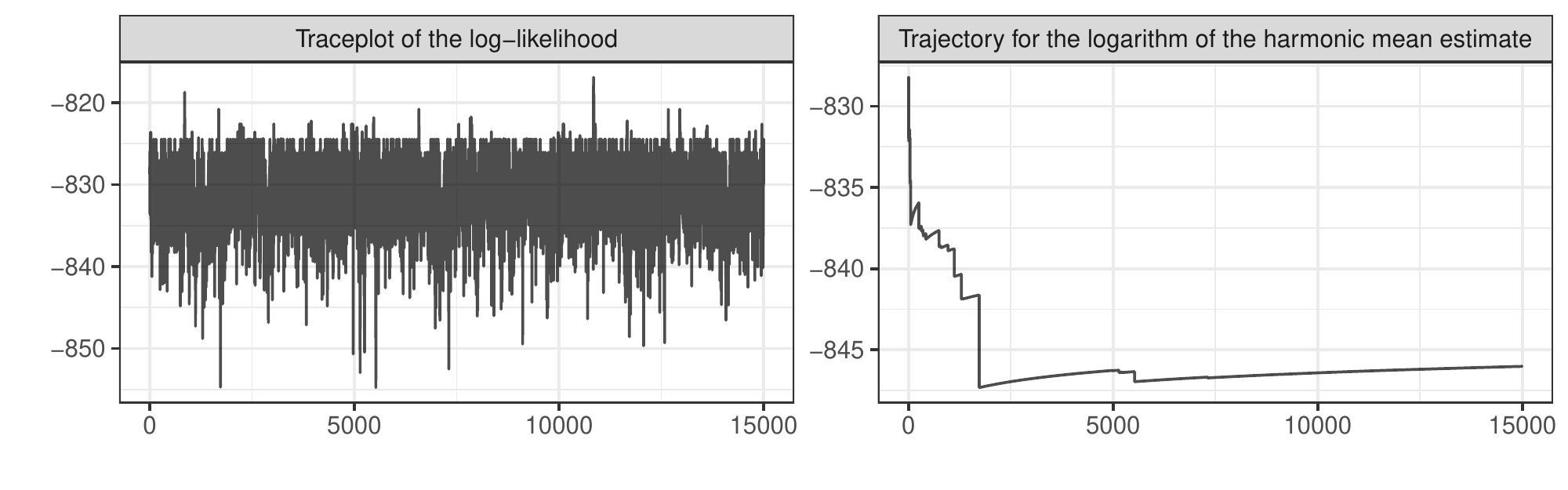}
\caption{\textsc{mcmc} diagnostics for the application. Left: traceplot for the logarithm of the likelihood in equation  \eqref{eq1} computed at the \textsc{mcmc} samples of $\bz$ after burn--in. Right: trajectory of the logarithm of the harmonic mean estimate in equation \eqref{harm} for growing $R$.}
\label{fig:2.1}      
\end{figure*}

Consistent with the above goal, we apply methods in Sections \ref{sec:1}--\ref{sec:2} to the $68 \times 68$ binary adjacency matrix $\bY$ encoding presence or absence of white matter fibers among anatomical regions in a representative Alzheimer's brain network provided by \citet{sulaimany_2017}. In this study,  brain regions are defined by the Desikan atlas \citep{desikan2006}, which provides additional information on hemisphere and lobe memberships \citep{kang2012}; see \citet{sulaimany_2017} for additional details on the construction of $\bY$. Figure \ref{fig:2} provides a graphical representation of $\bY$ with brain regions suitably reordered and organized in blocks according to the estimated endogenous assignments $\hat{\bz}$. The latter are obtained by considering the same \textsc{mcmc} settings and hyper--parameters of the simulation study, which proved effective and robust also in this application; see Figure \ref{fig:2.1}. As shown in Figure \ref{fig:2}, we learn $\hat{H}=12$ endogenous groups equally divided between the two hemispheres and showing an overall coherence of the partition structure across left and right regions. As expected, there is an evident  block--connectivity within hemispheres, although some groups  also display a  tendency to connect across hemispheres. For example, brain regions in the frontal lobe tend to create two highly interconnected clusters, one in each hemisphere, with these two blocks showing also a preference to create bridges among the two hemispheres. Despite these anatomical homophily structures, as highlighted in Figure~\ref{fig:2} and in Table~\ref{tab:2}, hemisphere and lobe partitions are not sufficient to fully characterize the endogenous block structures in Alzheimer's brains. There are, in fact, various sub--blocks within each hemisphere and these clusters typically comprise regions in different lobes.

\begin{table*}
\caption{Results of our proposed procedure for testing to what extent exogenous brain partitions are as effective as model $\mathcal{M}$ in characterizing the endogenous block structure for a  representative brain network of Alzheimer's individuals. Here, we focus on anatomical partitions and on grouping structures identified in representative brains of individuals characterized by three ordered cognitive decline stages. The \textsc{vi} distance between the estimated partition $\hat{\bz}$ under the infinite relational model  and the exogenous ones is also displayed.}
\label{tab:2}
\begin{tabular}{lccccc}
& \multicolumn{2}{c}{\textsc{Anatomical}} & \multicolumn{3}{c}{\textsc{Cognitive Decline Progression}}  \\
\noalign{\smallskip}
\hline\noalign{\smallskip}
$\bz^*$  & Hemispheres & Lobes &  \quad  Normal Aging & Early Decline & Late Decline   \\
\noalign{\smallskip}\hline\noalign{\smallskip}
$2 \log \hat{\mathcal{B}}_{\mathcal{M},\mathcal{M}^*}$ & $713.57$ & $1291.74$ & $156.25$ & $101.45$ & $41.12$ \\
$\textsc{vi}(\hat{\bz},\bz^*)$ & $2.29$ & $3.37$ & $1.46$ & $1.33$ & $1.10$  \\
\noalign{\smallskip}\hline
\end{tabular}
\end{table*}

We conclude by studying if the clustering structures inferred from representative brains of individuals in three ordered stages of cognitive decline can effectively explain the endogenous block structures in Alzheimer's brains. To accomplish this goal, we first apply Algorithm \ref{Algorithm} to the representative adjacency matrices of individuals characterized by normal aging, early and late cognitive decline  \citep{sulaimany_2017}, and then quantify, via the Bayes factors in Table~\ref{tab:2}, whether these partitions are  also effective in modeling the block structures within the Alzheimer's brain. Although $2 \log \hat{\mathcal{B}}_{\mathcal{M},\mathcal{M}^*}$ is above the threshold in  \citet{kass1995} suggesting strong evidence against this hypothesis for all three stages, it is interesting to notice how $2 \log \hat{\mathcal{B}}_{\mathcal{M},\mathcal{M}^*}$ decreases as cognitive decline progresses towards Alzheimers' disease. This means that the inferred partitions could be used, with caution, as a diagnostic strategy to identify the progress of the disease. The \textsc{vi} distances between $\hat{\bz}$ and these external partitions confirm the evidence provided by the Bayes factors, thus providing further support to our conclusions. 

To further validate the suitability of $\mathcal{M}$ as a flexible model for $\bY$, we also compute the in--sample missclassification error when predicting each $y_{vu}$ with $\hat{\theta}_{\hat{z}_v,\hat{z}_u}$. Such a measure is $0.1$, thus confirming that $\mathcal{M}$ can be regarded as a suitable model for this application.

\section{Discussion and Future Developments}
\label{sec:5}
In this contribution we propose a formal Bayesian testing procedure to assess the ability of a fixed exogenous node partition to characterize a network structure, with respect to an infinite relational model.  To accomplish this goal we compare an harmonic mean estimate of the marginal likelihood under this latter representation with the one induced by a stochastic block model conditioned on the external partition of interest. 
From a computational perspective, we rely on a collapsed Gibbs sampler which additionally allows Bayesian inference and uncertainty quantification on endogenous partitions. As illustrated in simulations and applications to brain networks, our proposal provides a simple yet effective procedure to obtain further insights regarding the effects of node attributes on network structures.

There are several  directions for future developments. For example, weighted networks comprising counts  or continuous measures of strength in the relationship can be easily incorporated within our strategy by simply replacing the  likelihood in \eqref{eq1} with a suitable one. This can be obtained by leveraging Poisson--gamma or Gaussian--Gaussian conjugacy, as done for the beta--binomial case. Moreover, while throughout the paper we have considered the problem of testing model $ \mathcal{M} $ against model $ \mathcal{M}^* $ given a single observed network $ \bY $, one may be interested in the same test given a sample of $ N $ exchangeable networks. This is feasible under our proposed framework and only requires to substitute $ p(\bY \mid \mathcal{\bz}) $ in \eqref{eq1} with $ p(\bY_1,\ldots,\bY_N \mid \bz)$. It is also possible to compare two exogenous partitions, rather than an exogenous and an endogenous one. This task is even simpler than the one analyzed in this article, since the likelihood in \eqref{eq1} can be computed in closed form for both the external partitions under comparison, thus avoiding the need of \textsc{mcmc} methods. For example, one may be interested in comparing an external assignment $\bz^*$ with a random permutation of the indices in such a vector to assess whether $\bz^*$ offers improvements in modeling network block structures or has no effect. Therefore, the perspective taken by \citet{bianconi2009} and \citet{peel2017} can be seen as a special case of our more general solution.

\begin{acknowledgements}
Sirio Legramanti and Daniele Durante acknowledge support from \textsc{miur--prin} 2017 grant 20177BRJXS. Tommaso Rigon acknowledges support from grant R01ES027498 of the National Institute of Environmental Health Sciences of the United States National Institutes of Health.
\end{acknowledgements}

\bibliographystyle{spbasic} 
\bibliography{EIRM}

\end{document}